\documentclass[%
 reprint,
superscriptaddress,
 amsmath,amssymb,
 aps,
]{revtex4-2}
\usepackage{graphicx}
\usepackage{dcolumn}
\usepackage{bm}
\usepackage{acronym} 
\usepackage{comment}
\usepackage{xcolor}
\usepackage{soul}
\usepackage{xspace}
\usepackage[colorlinks]{hyperref}
\usepackage[caption=false]{subfig}
\usepackage[export]{adjustbox}

\usepackage{float}

\newcommand{\astreos}{\texttt{ASTREOS}\xspace}

\begin{document}

\title{Rapid neutron star equation of state inference with Normalising Flows}

\author{Jordan McGinn}
\affiliation{%
SUPA, School of Physics and Astronomy, University of Glasgow, Glasgow G12 8QQ, United Kingdom
}%
\author{Arunava Mukherjee}
\email{arunava.mukherjee@saha.ac.in}%
\affiliation{
Saha Institute of Nuclear Physics, A CI of HBNI, 1/AF Bidhannagar, Kolkata-700064, India
}%
\affiliation{%
SUPA, School of Physics and Astronomy, University of Glasgow, Glasgow G12 8QQ, United Kingdom
}%
\author{Jessica Irwin}
\email{j.irwin.1@research.gla.ac.uk}
\affiliation{%
SUPA, School of Physics and Astronomy, University of Glasgow, Glasgow G12 8QQ, United Kingdom
}%
\author{Christopher Messenger}
\affiliation{%
SUPA, School of Physics and Astronomy, University of Glasgow, Glasgow G12 8QQ, United Kingdom
}%
\author{Michael J. Williams}
\affiliation{%
SUPA, School of Physics and Astronomy, University of Glasgow, Glasgow G12 8QQ, United Kingdom
}%
\author{Ik Siong Heng}
\affiliation{%
SUPA, School of Physics and Astronomy, University of Glasgow, Glasgow G12 8QQ, United Kingdom
}%

\date{\today}

\begin{abstract}
The first direct detection of gravitational waves from binary neutron stars on
the 17th of August, 2017, (GW170817) heralded the arrival of a new messenger
for probing neutron star astrophysics and provided the first constraints on
neutron star equation of state from gravitational wave observations.
Significant computational effort was expended to obtain these first results
and therefore, as observations of binary neutron star coalescence become more
routine in the coming observing runs, there is a need to improve
the analysis speed and flexibility. Here, we present a rapid approach for
inferring the neutron star equation of state based on Normalising Flows. As a
demonstration, using the same input data, our approach, \astreos, produces
results consistent with those presented by the LIGO-Virgo collaboration but
requires $< 1$~sec to generate neutron star equation of state confidence
intervals. Furthermore, \astreos allows for non-parametric equation of state
inference. This rapid analysis will not only facilitate neutron star equation
of state studies but can potentially enhance future alerts for electromagnetic
follow-up observations of binary neutron star mergers.  \end{abstract}

\maketitle


\acrodef{EOS}{equation of state}
\acrodef{ML}{machine learning}
\acrodef{NF}{normalising flow}
\acrodef{NS}{neutron star}
\acrodef{BNS}{binary neutron star}
\acrodef{BBH}{binary black hole}
\acrodef{NSBH}{neutron star-black hole}
\acrodef{GW}{gravitational wave}
\acrodef{PCA}{principle component analysis}
\acrodef{PE}{parameter estimation}
\acrodef{TOV}{Tolman-Oppenheimer-Volkoff}
\acrodef{p-p}{probability-probability}
\acrodef{EM}{Electromagnetic}

%
\emph{Introduction.}--- On 17 August 2017, during the second advanced detector
observing run, advanced LIGO~\cite{LIGOScientific:2014pky} and advanced
Virgo~\cite{VIRGO:2014yos} observatories detected the first \ac{GW} signal from
the coalescence of two \acp{NS}~\cite{GW170817,GW170817_MMA}. The global
network made a second \ac{GW} observation consistent with a signal from a
\ac{BNS} collision~\cite{GW190425} during the 3rd observing run. In addition, two
other \ac{GW} events likely originating from the mergers between pairs of
\acp{NS} and black holes were reported~\cite{NSBH}. To date, there have 
been more than 90 definitive \ac{GW} event detections, with the vast
majority being the merger of \acp{BBH}~\cite{gwtc3}. However, for merging
systems containing one or more \acp{NS} the presence of matter requires that
tidal effects be taken into account when modelling the inspiral stage of the
waveform. It is the tidal deformation that each star's gravitational field
induces on its companion that accelerates the decay of the inspiraling orbit
and imprints itself on the emitted \ac{GW} signals.  These detections therefore
provide a new opportunity to probe matter in extreme
conditions~\cite{LattimerPrakash_Science2004} present in the interior of these
stars, now confirmed through observation~\cite{GW170817,eos_gw170817}. 

%
\acp{NS} are the densest observable objects in the universe where densities
at the centre of the star can reach about $\sim$6--8 times the
saturation density ($\rho_{0}$)~\cite{LattimerPrakash_PhysRep2007, FBZRMD}.
Properties of hadronic matter at such a high density and low temperature have
not been probed by any experiment or observation except those targetting
\acp{NS}. A \ac{NS}'s tidal deformability parameter ($\Lambda$) along with its
mass ($m$), provides information on its interior composition via the underlying
\ac{EOS}~\cite{De_etal_prl_2018,Dietrich_etal_science2020,
ImamAM_etal_arXiv2305.11007}. Thus, astrophysical inference of $m$ and
$\Lambda$ of a \ac{NS} can provide useful constraints on the \ac{NS} \ac{EOS}.
This in turn provides valuable insights into the nature of nuclear interactions
of hadronic matter, which are governed by strong nuclear interactions at high
density/chemical potentials in degenerate
conditions~\cite{LattimerPrakash_ARNPS2021}. Each theoretical model for this
nuclear interaction provides a different \ac{EOS}, i.e., density ($\rho$) vs
pressure ($P$) relationship. 

%
The \ac{EOS} is assumed to be universal for all \acp{NS} in the sense that
there is a single relationship between $P$ and $\rho$ common for all such stars.
Each star is free to have its own independent mass $m$ and associated tidal
deformability $\Lambda$, governed by the underlying \ac{EOS}. A fully Bayesian 
approach~\cite{DelPozzo_bayesEOS2013} showed that detections of
$\mathcal{O}(10)$ \ac{BNS} mergers in the advanced detector era could constrain the
\ac{EOS} significantly. This work was expanded when~\cite{MCMC_Lackey_2015}
used Markov chain Monte Carlo simulations of the $P(\rho)$ relationships under a
piece-wise polytropic parametrisation of the \ac{EOS}~\cite{Read_etal_ppEOS}.
The first detection of \ac{BNS} event~\cite{GW170817} has allowed constraints
to be placed on the $m-\Lambda$ parameter plane. Follow-up analysis of the 
GW170817 event has been able to further constrain the \ac{EOS}~\cite{eos_gw170817,
AlMamun_etal_prl2021}. Subsequently, several Bayesian analyses have 
been performed to simulate inference for multiple \ac{BNS} events
\cite{Legred_etal_prd_2021,Huth_etal_nature_2022,Golomb_2022,Ray_etal_prd_2023}. 

The LIGO and Virgo collaborations applied two different methods of analysis to
inferring the \ac{EOS} of the \ac{BNS} merger GW170817~\cite{eos_gw170817}. In
one method, both components of the merger were assumed to be governed by the
same \ac{EOS}, represented by a spectral parametrisation to describe their
collective $P(\rho)$ relationship~\cite{Lindblom_spectralEOS}. The second
method is regarded as \ac{EOS}-insensitive but retains the assumption that both
\acp{NS} are governed by the same \ac{EOS}. Macroscopic properties of the
\acp{NS} are related to derive an expression for the asymmetric tidal
deformability of the components in a binary system. The relationships
established depend loosely on the \ac{EOS} and are tuned to a wide variety of
possible \ac{EOS} models, such that the individual tidal deformability pairs
recovered can map to the correct waveform template, independent of \ac{EOS}
model~\cite{Chatziioannou_etal_prd_2018}. Recently, there has been some effort
to incorporate non-parameteric inference of models using Gaussian processes 
conditioned on realistic \ac{NS}
\acp{EOS}~\cite{LandryAndEssick_nonparametric_eos_prd2019,
Essick_etal_nonparametric_eos_prd2020}.

%
The success of traditional Bayesian sampling techniques when applied to \ac{NS}
\ac{EOS} inference depends not only on the observed \ac{GW} signals but also on
the choice of parametrization. The simplest and most widely used form of the
\ac{NS} \ac{EOS} is the piecewise polytropic model~\cite{Read_etal_ppEOS}.
However, this representation is not unique, and it does not ensure causality of
sound speed for \ac{NS} matter. Moreover, applying traditional Bayesian
sampling techniques causes undesirable sampling issues~\cite{Carney_etal_prd_2018}
around the transition densities of different polytropic regions~\cite{MCMC_Lackey_2015}.
The spectral representation~\cite{Lindblom_spectral_repn, LindblomIndik_prd_2014}
provides us with a better parameterization of the \ac{EOS} that ensures both
thermodynamic stability and causality conditions. However, an implementation of
spectral parameterization for Bayesian sampling is computationally expensive
and cumbersome. Furthermore, it quickly becomes quite complex, for example, to
implement density-dependent constraints on the \ac{EOS} parameter space
corresponding to different terrestrial nuclear physics experiments at lower
densities, along with astronomical observations at the medium density regime,
and model agnostic and non-parametric aspects of the higher density
part~\cite{FBZRMD, Naresh_withAMandBKA_prc2023} of the \ac{EOS}. 

Thus, the existing traditional Bayesian sampling methods have a major limitation in 
performing statistical inference from different families of models corresponding
to different sets (and also different numbers) of parameters. This is in
addition to the fact that such techniques can also be extremely computationally
expensive. The issue of analysis latency is further compounded by the
significant increase in the number of \ac{BNS} detections expected as the
sensitivities of advanced detectors improve~\cite{prospects}. To perform \ac{EOS}
inference on \ac{BNS} merger events during future \ac{GW} advanced detector
observing runs, rapid inference tools are needed to act upon event data at low
latency that can be easily combined with other standard low-latency pipelines.

%
Recent advances in \ac{ML} have been successfully applied to \ac{GW} data
analysis~\cite{Cuoco_2020}. These include parameter estimation
\cite{Gabbard_etal_NatPh2022, chua_2020, green_2020, williams_2021, 
Dax_etal_PRL2021, Dax_etal_PRL2023}, waveform modelling~\cite{McGinn_2021, willaims_gpr}
and searches~\cite{2017arXiv171107966G, gabbard_cnn, bayley2020}. One clear
motivation behind using \ac{ML} to either supplement or replace existing algorithms
is that, once trained, the \ac{ML} algorithm can run using a significantly lower fraction
of the computational cost and time. \ac{ML} algorithms are also naturally more
flexible and can account for hard-to-model or un-modelled components of the
analysis.

%
In this letter, we describe an \ac{ML} approach to provide posterior distributions
describing the \ac{NS} \ac{EOS}. As input we use the standard parameter
estimation data products (posterior samples) on the gravitational masses ($m$) and
the tidal deformabilities ($\Lambda$) of the components within a \ac{BNS} system. 
Such data are generated as a standard output product as part of the data release
associated with detected \ac{BNS} events~\cite{data_release}. The
output of our \ac{NF} \cite{nfreview1,nfreview2} based analysis is an ensemble
of non-parametric \acp{EOS} describing the relationship between pressure and
energy density within \acp{NS}. In addition, we output correlated samples of
\ac{NS} central densities and maximum permitted densities corresponding to its
TOV-limit. Hence, once trained, the \ac{NF} acts as a rapid functional generator 
of plausible \ac{NS} \acp{EOS}.

%
\emph{Method.}--- A \ac{NF} is a generative \ac{ML} model
that learns to transform samples from a given distribution to a simpler
``latent'' distribution via a series of invertible mappings.
One advantage \acp{NF} have over other generative models (Generative Adversarial
Networks~\cite{Goodfellow2014}, Variational Auto-Encoders~\cite{VAE2014}) is that
they learn the probability density function of the training data explicitly.
At the training stage, a conditional \ac{NF} learns to map samples $x$ in the
data space $\mathcal{X}$ and $y$ in the conditional data space $\mathcal{Y}$ to
a point $z$ in latent space $\mathcal{Z}$ such that the following function
holds:
\begin{align}
p_{\mathcal{X} | \mathcal{Y}}(x | y) = p_{\mathcal{Z}}(f(x|y))~\bigg | \textrm{det} \bigg ( \frac{\partial f(x|y)}{\partial x} \bigg ) \bigg |,
\label{eq:normflow}
\end{align}
where $f$ is the bijective mapping $f: \mathcal{X} \rightarrow \mathcal{Z}$ and
$\partial f(x|y) / \partial x$ is the Jacobian of $f$ evaluated at $x$. 
This requires transforms whose Jacobian determinants are tractable and
easy to compute. In this work we use a real non-volume preserving (Real 
NVP)~\ac{NF}~\cite{dinh2017density} which uses stacked
\textit{coupling transforms}. As the learned transforms are invertible, samples
can be drawn from $p_{\mathcal{X}|\mathcal{Y}}$ by sampling from
$p_{\mathcal{Z}}$ and applying the inverse transform $f^{-1}$.
(For further details on \acp{NF}, please see~\cite{nfreview1,nfreview2}.)

%
Our aim is to use a conditional \ac{NF} to accurately approximate a function that
maps values of the component masses and tidal deformabilities of a \ac{BNS}
system, $y = (m_1, m_2, \Lambda_1, \Lambda_2)$, to an associated set of
\ac{EOS} information $x$, parameterised for convenience by a \ac{PCA}
decomposition (see~\ref{preprocessing_data_training} for details), the central
densities of each star $\rho_{\text{c}}$, and a maximally allowed central
density $\rho_{\text{max}}$ for the \ac{EOS} in question. We refer to these
latter three quantities as \textit{auxiliary parameters} and we call our analysis
\astreos.

%
Our training data consists of $10^5$ phenomenological \ac{NS} \acp{EOS}
represented initially as energy-density computed on a fixed vector of pressure
values. This is compressed into a lower-dimensional representation using
\ac{PCA} and accompanied by the maximum allowed energy density
$\rho_{\text{max}}$ for each \ac{EOS} - this and the auxiliary parameters define
the data space $\mathcal{X}$.
To associate each $x$ to a corresponding \ac{GW} observable $y$ we first sample
from a uniform prior on the joint mass distribution with fixed lower bound
$0.5\text{M}_{\odot}$, an upper bound dictated by the \ac{EOS} and
$\rho_{\text{max}}$, and the constraint $m_1 \ge m_2$. These choices then define
the value of tidal deformability for each \ac{NS} in the binary as well as the
central energy-density of each component.    

%
We use an implementation of \acp{NF} called \texttt{GLASFLOW}~\cite{glasflows}
based on \texttt{NFLOWS}~\cite{nflows} which is written for
\texttt{PYTORCH}~\cite{NEURIPS2019_9015}. Our present version is 
implemented by a modification of an initial network outlined in~\cite{nseos_ml}. 
We train the flow for 5000 epochs
with a batch size of 4096 and initial learning rate of 0.005 which is
decayed to zero using cosine annealing~\cite{cosine_annealing} together 
with the AdamW~\cite{AdamW} optimiser. There are two residual blocks for each of the three
coupling transforms which each contain 151 neurons. We use Batch Normalisation between
each coupling transform~\cite{dinh2017density} and optimise
hyper-parameters using ``Weights \& Biases''~\cite{wandb}. Data is
split between 80\% training and 20\% validation and on a NVIDIA Tesla V100 (or
equivalent) training requires $\mathcal{O}(3)$ hours with a GPU memory
footprint of $\sim 2$GB.

%
The \ac{NF} allows us to model the distribution
$p_{\mathcal{X}|\mathcal{Y}}(x|y)$ but our knowledge of $y$ for a given
\ac{BNS} event is limited to the
posterior $p(y|h)$ given the observed strain $h$. Therefore we must marginalise over
the correlated uncertainties in $y$ due to the \ac{GW} detector noise and other
correlations between these and other \ac{GW} parameters. Our final
inference on the \ac{EOS} and auxilliary parameters for a given \ac{BNS}
observation is given by
\begin{align}\label{eq:marg_post}
    p(x|h) = \int p_{\mathcal{X}|\mathcal{Y}}(x|y)p(y|h)\,dy
\end{align}
where, rather than evaluating this integral directly, samples from $p(x|h)$ are
obtained by drawing an equal number of samples (usually one) from
$p_{\mathcal{X}|\mathcal{Y}}(x|y_j)$ for each posterior sample $y_j$.

%
\begin{figure}
    \centering
    \includegraphics[width=\columnwidth]{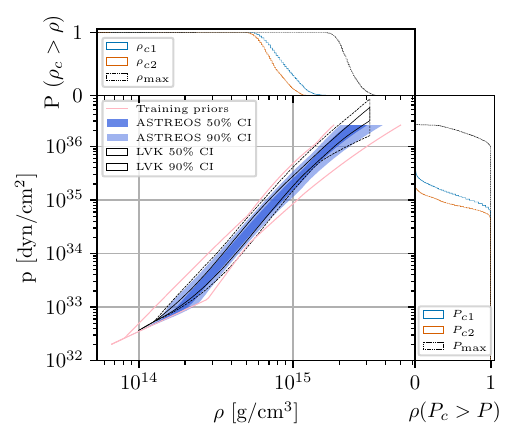}
    \caption{The pressure $p$ as a function of energy-density $\rho$ for the
GW170817 event. We show posteriors on energy density as a function of pressure
obtained from \astreos (blue bands) together with posteriors taken from
GW170817 LIGO-Virgo spectral parameterisation analysis~\cite{eos_gw170817} (black
lines) and the \astreos prior training bounds (pink lines).
The dark blue and light blue shaded regions for \astreos, and the solid and
dashed lines for the LIGO-Virgo analysis, correspond to the symmetric $90\%$
and $50\%$ confidence intervals respectfully. The top panel shows the \astreos
posterior survival functions on the central energy-densities of each star
$\rho_{\text{c},1}$, $\rho_{\text{c}c,2}$ (blue and red), and the maximum
energy-density $\rho_{\text{max}}$ (black). The panel on the right shows the
same for the pressure quantities $p_{\text{c},1}$, $p_{\text{c},2}$ (blue and
red) and $p_{\text{max}}$ (black).}\label{fig:gw170817-compare} 
\end{figure}


\emph{Results.}--- We demonstrate the effectiveness of our model by analysing
the GW170817 event using posterior
samples~\footnote{https://dcc.ligo.org/LIGO-P1800370/public} associated
with~\cite{gwtc1} available from the Gravitational Wave Open
Science Center~\cite{GWOSC}. These samples do not assume any inherent
correlation between individual \ac{NS} tidal deformabilities or that they share
the same universal \ac{EOS}. 

To perform \ac{EOS} inference of GW170817 using \astreos, as input we used the
joint posterior samples of $m_1,m_2,\Lambda_1,\Lambda_2$ to represent the
conditional $y$ component in the \ac{NF}. These posterior samples were computed
assuming low-spin priors on the point-mass parameters, the most relevant of which 
are the uniform priors on the detector frame component mass with constraints
$0.5\text{M}_{\odot} \leq m_2 \leq m_1 \leq 7.7\text{M}_{\odot}$ and uniform
priors on component tidal deformabilities $\Lambda_{1,2}\in [0,5000]$. 

In order to use the \ac{NF} model for testing, we draw one latent space sample
from $p_{\mathcal{Z}}$ for each posterior sample $y_j$ from the GW170817
posterior $p(y|h)$ and apply the \ac{NF} mapping to obtain a sample $x$. We discard 
any posterior samples that lie outside of our prior training space. We repeat the
process according to Eq.~\ref{eq:marg_post} to build up a distribution of samples
from the final \ac{EOS} posterior $p(x|h)$, discarding any output samples
that lie outside our 10-dimensional prior training space~\footnote{We use a
combination of Convex Hull modelling and a Gaussian mixture model of the prior
to perform sample rejection for the conditional $y$ and \ac{EOS} $x$ space
respectively.}. It takes only $\mathcal{O}(0.1)$~sec to generate and convert
2500 posterior samples to correctly normalised \ac{EOS} curves. 

Our main GW170817 result is compared to the spectral parametrisation analysis
of~\cite{eos_gw170817} in Fig.~\ref{fig:gw170817-compare} where we show
\ac{EOS} confidence intervals on energy-density as a function of pressure. We
also show cumulative probability curves obtained from \astreos for pressure and
energy-density for each \ac{NS} and similar curves for the corresponding
maximum allowed pressures and energy-densities of the inferred \ac{EOS}. We
note that a possible reason for the discrepancy between the results towards
higher energy densities is the difference between the training prior
distribution of \acp{EOS} and that assumed for the LIGO-Virgo analysis. Both
the LIGO-Virgo spectral paramaterisation and the input samples to \astreos use
priors uniform in component mass. We can also rule out the choice of the lower
component mass prior boundary used in training the \ac{NF} (since they are
identical) and although the upper bounds do differ (the maximum component mass
used in training is $3\text{M}_{\odot}$) we retain only GW170817 samples lie within the
training space.

\begin{figure} 
\includegraphics[width=1\columnwidth]{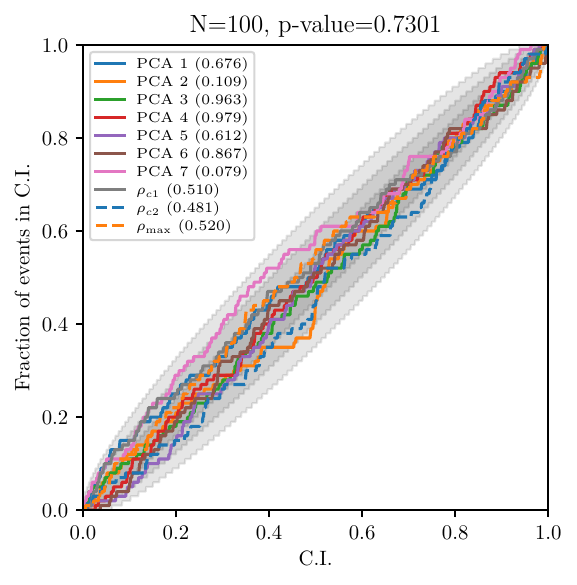}
\caption{A \ac{p-p} plot on the 10 inferred \ac{EOS} and auxilliary parameters
output from the \astreos analysis using 100 instances of conditional data $y$
drawn from the validation set. The bracketed quantities in the plot legend
refer to the p-values of each parameter curve under the null hypothesis that
each curve traces the diagonal. The gray error bands represent the 1,2 and
3$\sigma$ confidence intervals for each curve and the combined parameter
p-value for the null hypothesis is 0.7301.}\label{fig:pp_plot}
\end{figure}

To further verify the statistical consistency of the \astreos analysis, we
provide a \ac{p-p} plot~\cite{Cook:2006pp,Talts:2018pp} in Fig.~\ref{fig:pp_plot} with curves
for each of the output \ac{EOS} and auxiliary parameters. It shows the fraction
of true parameter values that lie within a given confidence interval as a
function of confidence interval. Curves that trace the diagonal within the
uncertainty bounds dictated by the number of simulated cases, indicate
statistical consistency of the inferred parameter distributions output from the
\ac{NF}. This result represents an isolated test of the statistical robustness of
the \astreos analysis and does not include the use of posterior \ac{GW} samples
$p(y|h)$. The process tests the statistical consistency of the $p(x|y)$ (the
\ac{NF} component) only. In this test we draw single instances of \ac{EOS} and
auxilliary parameters ($x$) from our validation set together with their
associated mass and tidal deformability parameters ($y$). For the purposes of
the \ac{p-p} test, $x$ then defines the ``truth'' and $y$ is then input to
\astreos to obtain 2000 samples drawn from the posterior $p(x|y)$. This process
is repeated 100 times allowing the construction of the \ac{p-p} plot shown in
Fig.~\ref{fig:pp_plot} and resulting in a combined p-value of 0.7301. There is
no additional marginalisation over the uncertainty on $y$ as would be the case
had we simulated samples from $p(y|h)$.  This allows us to verify the
correctness of the \ac{NF} unaffected by additional independent components of a
full \ac{GW} analysis pipeline.        

%

%
\emph{Conclusions.}--- In this Letter we have for the first time demonstrated
that neural networks can accurately infer \ac{NS} \ac{EOS} curves conditioned
on prior measurements of component masses and tidal deformabilities within a
strict Bayesian framework. The \ac{NF} architecture of our analysis learns the
distribution of plausible \acp{EOS} consistent with the conditional argument
and naturally marginalises over the uncertainty inherent to the \ac{GW}
measurement. The analysis also provides estimates of the central
energy-densities of both the component \acp{NS} and the maximally allowed
energy-density of the \ac{EOS}. 

Once trained, the network can generate $\sim$25000 samples per second from
the \ac{EOS} posterior meaning that \ac{EOS} posteriors can be obtained almost
immediately after a \ac{GW} \ac{PE} analysis has completed. We note that the
\ac{NF} need only be trained once, after which all events can be analysed
rapidly. Coupled with the growing number of ultra-low latency \ac{ML} \ac{PE}
pipelines~\cite{Gabbard_etal_NatPh2022, green_2020,Dax_etal_PRL2023}, rapid
inference on \ac{EOS} properties can then be used to inform low-latency \ac{EM}
follow-up observations regarding the likelihood of prompt \ac{EM} emission. 

We would like to stress the model independence of our approach and clarify that
although the \ac{EOS} outputs of our analysis are parameterised via \ac{PCA}
coefficients, the training data can in principle be composed of any \ac{EOS}
model or mixture of models. It is therefore the choice of training data that
defines the prior on the \ac{EOS} parameter space and allows the user to be as
constraining, or flexible, as they want. The minimal requirement from any model is
that it is possible to sample a component mass and compute the corresponding tidal
deformability for a given \ac{EOS} represented via a vector of pressure as a
function of energy-density. We would also note that our analysis can be
easily modified to handle \ac{NSBH} systems~\cite{NSBH, gwtc3} where the
conditional data in this case would consist of only one pair of mass and tidal
deformability parameters. 

As is evident from the GW170817 result presented in
Fig.~\ref{fig:gw170817-compare}, for individual \ac{GW} events the information
content and constraints placed on the prior \ac{EOS} space is relatively
modest. However, the number of additional \ac{BNS} (and \ac{NSBH}) expected
events~\cite{2018LRR....21....3A} in the ongoing O4 observing run
of the advanced \ac{GW} detector network will allow \ac{EOS} inference results
to be combined. With this combination on the universal \ac{EOS} parameters we
would expect an approximate $\sqrt{N}$ reduction~\footnote{This rough scaling
behaviour is only really true for large $N$ where the variation in individual
event signal-to-noise ratio is averaged over. GW170817 was a relatively loud
event and hence the initial trend with increasing detections is likely to be
less powerful than the $\sqrt{N}$ relation.} in the uncertainty represented in
Fig.~\ref{fig:gw170817-compare} where $N$ is the number of detections.
Fortunately \acp{NF} are ideally suited for combining results by making use of
the trained Jacobian (see Eq.~\ref{eq:normflow}). Instead of generating samples
from $p(x|h)$, it is possible to directly evaluate the distribution for each
event either on a common grid of $x$ values or through a sampling
algorithm, e.g., Markov chain Monte Carlo.       

\emph{Acknowledgement.}--- This material is based upon work supported by NSF's
LIGO Laboratory which is a major facility fully funded by the National Science
Foundation. The authors would like to thank members of the LIGO-Virgo-Kagra
extreme matter group for valuable input, specifically Jocelyn Read,  Lami
Suleiman, Katerina Chatziioannou, Isaac Legred, Jolien Creighton, and 
Md. Emanuel Hoque. AM acknowledges support from the
DST-SERB Start-up Research Grant SRG/2020/001290 and partial
computational support from Newton-Bhabha Grant. AM, CM, ISH acknowledge
partial support from Science and Technology Research Council (Grant No.
ST/L000946/1). CM is also supported by the European Cooperation in Science and
Technology (COST) action [CA17137].

The authors acknowledge the usage of the analysis software
\texttt{NFLOWS}~\cite{nflows}, \texttt{GLASFLOW}~\cite{glasflows},
\texttt{BILBY}~\cite{bilby_2019}
and open data resource from GWOSC~\cite{GWOSC_OpenData_2021SoftX}. 

\bibliography{nseos_draft.bib}

\section{Supplemental Material}

%
\subsection{Pre-processing and data-preperation for training}\label{preprocessing_data_training}

Within our training set, each \ac{EOS} consists of energy density pre-computed
on a fixed grid of 256 pressure values which is then truncated to include the
105 grid points spanning the pressure range from $2.024\times 10^{32}$
dyn/cm$^{2}$ to $2.517\times 10^{36}$ dyn/cm$^{2}$. This preprocessing step is
motivated by the necessity to learn only the high density regions of the
\ac{EOS} parameter space where the prior is not already well constrained. For
each \ac{EOS} there exists a different range of possible neutron star masses.
We define a uniform prior on this range with a lower bound of 0.5M$_{\odot}$ for
all \acp{EOS} up to the maximum possible mass allowed by each \ac{EOS}. 

We note that although this is a specific choice of component mass prior it does
not directly influence the output of our \ac{NF}. As is clear from
Eqs.~\ref{eq:normflow}~\&~\ref{eq:marg_post}, the posterior probability on the
\ac{EOS} and auxilliary parameters ($x$) modelled by the Flow is conditional on
the masses and tidal deformabilities ($y$). The distribution of $y$ is
determined soley by the \ac{GW} posterior $p(y|h)$ (where specific and
potentially influential mass and tidal deformability priors have already been
applied) and not influenced by their prior distribution during training.
The only practical consequence of the choice of training prior is that we should
not trust the \ac{NF} to be well behaved when used on mass and tidal
deformability values outside the prior training boundaries. Assuming that these
boundaries are well motivated, we can therefore discard any violating input
samples from the \ac{GW} posterior $p(y|h)$ before an analysis.           

%
When selecting training data samples for input to the \ac{NF} we choose the
\ac{EOS} uniformly from the $10^5$ training data examples. This then defines
the component mass prior range allowing us to sample the component masses $m_1$
and $m_2$ whilst ensuring that $m_1 \geq m_2$. The \ac{EOS} and masses then
naturally determine the central densities of each star in the binary, the tidal
deformabilities, and the maximum density allowed by the \ac{EOS}. With this
scheme, it is possible to reuse each of the training \acp{EOS} to augment our
training data with different choices of component mass. This would allow the
Flow to see examples of different binary confgurations for repeated \acp{EOS}
and therefore help the model to generalise over the training prior.   

%
In order to reduce the dimensionality of the problem, ensuring stable and fast
training, and to provide a general compression scheme applicable to all
\ac{EOS} model choices, we use a \ac{PCA} representation~\cite{Jolliffe} of the
density as a function of pressure. We choose to represent the \ac{EOS} data
using 7 PCA components (accounting for 99.975\% of the variation in the prior
training set). All of our training \ac{EOS} data can therefore be represented
with high fidelity using a linear combination of our 7 principle eigenvectors
(each of length 105 samples). The data space $\mathcal{X}$ is therefore reduced
to a total of 10 dimensions (7 \ac{PCA} components plus the 3 auxiliary
parameters). As a final preprocessing step, \ac{EOS}, auxiliary, and
conditional parameters are standardised separately by subtracting the mean and
scaling to have unit variance. Further, to account for the large dynamic range
of the tidal deformability values and central and maximum densities, they are
represented by their natural logarithms before standardisation and input to
training.

%
\subsection{Equation of State data used for training the network}\label{EOS_data}

We simulate $10^{5}$ phenomenological neutron
star \acp{EOS} to train the Flow model. To accommodate the analysis within
limited computational resources, we generated the \acp{EOS} from a 3-piece
polytropic neutron star \ac{EOS}-family widely used in the literature
\cite{Read_etal_ppEOS}. Each \ac{EOS} contains a low-density crust described
by the SLy4 \ac{EOS}~\cite{SLy4} but at higher densities behaves as a piece-wise
polytrope with transition densities at $5\times 10^{14}$ g/cm$^{3}$ and $10^{15}$
g/cm$^{3}$. We empirically choose the polytropic indices and their distributions
in such a way that the variation in our \ac{EOS} training set closely follows
the prior data-set used in~\cite{GW170817}. As we discuss in the conclusions,
our method allows for flexibility in the choice of prior \ac{EOS} model, its
parameterisation, and the prior distribution of its parameters. We remind the
reader that in any Bayesian analysis the final posterior will to some degree
be influenced by the choice of prior.

%
\subsection{Technical detail about the training method}\label{training_method}

The training of \astreos and indeed most \acp{NF} involves the process
of learning a forward mapping from the training data (the \ac{EOS} and
auxiliary parameters) conditional on labels (the \ac{NS} component masses and
tidal deformabilities) to a zero-mean unit-variance uncorrelated
multi-dimensional Gaussian with total dimension equalling that of the training data.
Training is performed with the goal of minimising the KL divergence between the 
learned distribution $p_{\mathcal{X}|\mathcal{Y}}(x|y)$ and the distribution of samples
in the data space $\mathcal{X}$.
Once trained, we can use \astreos to perform the inverse mapping from a single
condition label $y$ and a randomly drawn location $z$ from the latent space
distribution, to an \ac{EOS} and corresponding auxiliary values $x$. As we can
continually draw random points from the latent space to produce \ac{EOS} data
using the same conditional labels, there will naturally be variation in the
output \ac{EOS} data. This is encoded in the Flow output distribution
$p_{\mathcal{X}|\mathcal{Y}}(x|y)$. The variation within this distribution is
representative of the degeneracy inherent within \ac{EOS} inference based on
single estimates of pairs of component masses and tidal deformabilities. A single value
of $y$ maps to a distribution of plausible \acp{EOS} all consistent with the
input conditional data and the prior distribution represented by the training
data. We repeat this process over a set of mass and tidal deformability samples
drawn from the joint posterior $p(y|h)$, where $h$ is the \ac{GW} strain data
for a particular \ac{BNS} event. By doing this we are able to marginalise over
the correlated uncertainties in $y$ due to the \ac{GW} detector noise and other
correlations between these and other \ac{GW} parameters via
Eq.~\ref{eq:marg_post}.

\end{document}